% Official macros for papers submitted to "Baltic Astronomy"
% First version, March 14, 1991, by Enn Saar (saar@aai.tartu.ew.su).
% Do not change this file yourself; send suggestions
% to the editors of the journal, who will update the macros.
%
% OVERALL SIZES
%
\magnification=\magstep1
\baselineskip=11pt plus .1pt minus .1pt
\hsize=12.5truecm
\vsize=19.0truecm  % with head--and footlines gives 20.5 cm
\hfuzz=5pt\vfuzz=5pt
\tolerance=1000
\overfullrule=0pt
\parskip=0pt
\abovedisplayskip=3 mm plus6pt minus 4pt
\belowdisplayskip=3 mm plus6pt minus 4pt
\abovedisplayshortskip=0mm plus6pt minus 2pt
\belowdisplayshortskip=2 mm plus4pt minus 4pt
\predisplaypenalty=0
\clubpenalty=10000
\widowpenalty=10000
\parindent=2em
%
% FONTS
%
\font\pgnumfont=cmr9
\font\headlinefont=cmti9
%\font\titlefont=cmbx10 scaled\magstep2
 \font\titlefont=cmbx10
%\font\titlefont=cmbx10
%\font\authorfont=cmr12
\font\authorfont=cmr10
%\font\addressfont=cmti10
\font\addressfont=cmti9
\font\datefont=cmr9
%\font\sumfont=cmr10
\font\sumfont=cmr9

\font\absfont=cmbx9
%\font\secfont=cmbx12
\font\secfont=cmr10
\font\subsecfont=cmti10
%\font\subsecfont=cmti12
\font\subsubsecfont=cmr10
%\font\subsubsecfont=cmr12
\font\figfont=cmr9
\font\figheadfont=cmbx9

\font\tabheadfont=cmbx9
\font\mainfont=cmr10
\font\petitrm=cmr9

%
%
% Header variables
%
\newtoks\TITLE \newtoks\AUTHOR \newtoks\ADDRESS \newtoks\SUMMARY
\newdimen\sumindent \sumindent=\parindent
\newtoks\KEYWORDS \newtoks\SUBMITTED \newtoks\ACCEPTED
\newtoks\SENDOFF
%

% HEADER LINES
%
% (for a full number the variables \VOLUME (common) and
% \PAGES (separately for every article) must be given:
% e.g., \VOLUME={1,} \PAGES={11--23.}
%
\newtoks\firstpage
\let\firstpage=Y
\newtoks\AUTHORHEAD \newtoks\ARTHEAD \newtoks\VOLUME \newtoks\PAGES
\if!\the\AUTHORHEAD!\AUTHORHEAD={\the\AUTHOR}\fi
\if!\the\ARTHEAD!\ARTHEAD={\the\TITLE}\fi
\footline={\hfil}
\headline={\ifodd\pageno\rightheadline \else\leftheadline\fi}
\def\leftheadline{\if Y\firstpage\firsthead\global\let\firstpage=N
  \else\lefthead\fi}
\def\rightheadline{\if Y\firstpage\firsthead\global\let\firstpage=N
  \else\righthead\fi}
\def\lefthead{\pgnumfont\number\pageno\hfil\headlinefont\the\AUTHORHEAD}
\def\righthead{\headlinefont\the\ARTHEAD\hfil\pgnumfont\number\pageno}
\def\firsthead{\headlinefont Baltic Astronomy,~vol.\the\VOLUME,
\the\PAGES,~\the\year .\hfil}
\voffset=2\baselineskip % Knut recommends this
%
% PRINT HEADER

\newdimen\oldbaselineskip \oldbaselineskip=\baselineskip
\def\test#1{\newlinechar=`@\if!\the#1! \message{#1 not given@}\fi}%
\def\printheader{
  \parindent=0pt
  %\ifodd\pageno \else\advance\pageno by1\fi%
  \null\vskip1.cm
  \test{\TITLE}
  \vbox{\baselineskip=15pt
    \titlefont\the\TITLE
    }
  \vskip8mm plus8mm
% \vskip6mm plus4mm
  \test{\AUTHOR}
  \authorfont\the\AUTHOR
% \vskip3mm
  \vskip2mm
  \test{\ADDRESS}
  \addressfont\the\ADDRESS
  \vskip2mm
  \test{\SUBMITTED}
  \line{\datefont Received \the\SUBMITTED
    \if!\the\ACCEPTED!\else, accepted \the\ACCEPTED\fi.\hfill}
  \vskip4mm plus4mm
  \vbox{\leftskip=\sumindent\parindent=0pt
    \parskip=5pt
    \absfont Abstract.
    \test{\SUMMARY}
    \sumfont\the\SUMMARY\par
    \absfont Key words:
    \test{\KEYWORDS}
    \sumfont\the\KEYWORDS\par
    }
  \sumfont
  \if!\the\SENDOFF!\else\footnote{}{Send offprint requests to:
 \the\SENDOFF}\fi
  \parindent=2em
  }
%
% SECTION HEADERS
%
\newdimen\uppergap \newdimen\lowergap
%\uppergap=7mm \lowergap=4mm
\uppergap=5mm \lowergap=3mm
\newdimen\secind \newdimen\subsecind \newdimen\subsubsecind
\setbox0=\hbox{\secfont 9. }\secind=\wd0
\setbox0=\hbox{\subsecfont 9.9. }\subsecind=\wd0
\setbox0=\hbox{\subsubsecfont 9.9.9. }\subsubsecind=\wd0
\def\section#1{\goodbreak\par\vskip\uppergap
  \noindent\hangindent\secind\hangafter=1\secfont#1
  \vskip\lowergap\mainfont\par\nobreak}
\def\subsection#1{\goodbreak\par\vskip\uppergap
  \noindent\hangindent\subsecind\hangafter=1\subsecfont#1
  \vskip\lowergap\mainfont\par\nobreak}
\def\subsubsection#1{\goodbreak\par\vskip\uppergap
  \noindent\hangindent\subsubsecind\hangafter=1\subsubsecfont#1
  \vskip\lowergap\mainfont\par\nobreak}
%
% FIGURES
%
\def\WFigure#1#2#3{\goodbreak\midinsert\vbox{%\parindent=0pt
  \null\centerline{#2}\vskip1.5truemm
  \figheadfont\indent Fig.~#1.\figfont\ #3
  \par\mainfont
  }\endinsert}
%

%

%

%

% TABLES
%
\newdimen\tabind
\setbox0=\hbox{\tabheadfont Table 55.} \tabind=\wd0

%
% REFERENCES
%
%\def\References{\par\noindent\vskip\uppergap
\def\References{\vskip\uppergap
\line{\secfont REFERENCES\hfill}
  \vskip0.8\lowergap
 \petitrm
% \bgroup\petit
  }
\def\ref{\goodbreak
\hangindent12pt\hangafter=1
\noindent\ignorespaces}
\def\endref{\egroup}
%
% FINAL COMMAND
%
% submitting individual papers
\def\byebye{\egroup\par\vfill\supereject\end}
% in a full journal
%\def\byebye{\egroup\par\vfill\supereject}
%
% PETIT -- for references
%

%
% USEFUL ASTRONOMICAL ABBREVIATIONS (from AA macros)
%

\def\utw{\smash{\rlap{\lower5pt\hbox{$\sim$}}}}
\def\udtw{\smash{\rlap{\lower6pt\hbox{$\approx$}}}}

%those are some definitions for some help making plane TeX tables.

%3.5mm
%3.5mm
  %1.75mm

\def\add{\vrule height0pt depth0pt width.5em}
\def\addu{\vrule height0pt depth0pt width1.3em}

   %all of them
 %not visible

\def\ddown{\lower2.5ex\hbox}
\def\ddow{\lower1.7ex\hbox}
\def\down{\lower1ex\hbox}
\def\uppp{\raise1ex\hbox}
\def\dnnn{\lower1ex\hbox}
\def\uuppp{\raise2ex\hbox}

\def\(o-c){$O-C$}

 %ApJlikes it this way now

\def\angstr{A\kern-.56em\raise1.9ex\hbox{$\scriptscriptstyle\circ$}$\,$}

\newdimen\free\newdimen\shift
\def\Entry#1#2#3{\par\goodbreak\smallskip%
  \setbox1=\vbox{\advance\hsize by-10mm\parindent=0pt
    \def\\{\par}%
    \it#1. \rm#2}
  \line{\box1\hfill#3}\smallskip
}%
% WIDE TABLES, MUST BE PRINTED AS A SEPARATE FILE!
\newdimen\savesize

\def\shiftfigure #1#2#3#4#5{
    \vbox to #2 { \ifodd #5 \rightskip#4 \else\leftskip#4 \fi
                  \null\vfil
                  \figheadfont Fig.~#1.\figfont #3
                  \medskip
                }
                          }

%%%%%
\year1999
%%%%%


\input psfig.sty

\TITLE={Spectral analysis of four multi mode pulsating sdB stars}
\VOLUME={8}
\PAGES={XXX--XXX}              
\pageno=1                      
\AUTHORHEAD={U. Heber, I.N. Reid, K. Werner}
\AUTHOR={Ulrich Heber$^{1}$, I. Neill Reid$^{2}$, Klaus Werner$^{3}$ }
\ARTHEAD={Spectral analysis of pulsating sdB stars}
\ADDRESS={
\item{}$^{\add 1}$ Dr. Remeis-Sternwarte Bamberg, Astronomisches Institut 
\item{}\addu der Universit\"at Erlangen-N\"urnberg, D-96049 Bamberg, Germany
\item{}$^{\add 1}$ Palomar Observatory, 
\item{}\addu Pasadena, USA
\item{}$^{\add 3}$ Institut f\"ur Astronomie und Astrophysik, Waldh\"auser
  Stra\ss e 64,
\item{}\addu D-72076 T\"ubingen, Germany
}
\SUMMARY={Four members of the new class of pulsating sdB stars are analysed
from Keck HIRES spectra using 
NLTE and LTE model atmospheres.  
Atmospheric parameters (T$_{\rm eff}$, log~g, log(He/H)), metal 
abundances and rotational velocities
are determined. Balmer line fitting is found to be consistent 
with the helium 
ionization equilibrium for PG$\,$1605+072 but not so 
for PG$\,$1219$+$534 indicating that 
systematic errors in the model atmosphere analysis of the latter 
have been underestimated previously. All stars are found to be helium deficient 
probably due to diffusion. The metals are also depleted with the notable 
exception of iron which is solar to within error limits in all four stars, 
confirming predictions from diffusion calculations of Charpinet et al. (1997). 
While three of them are slow rotator's (v$\,$sini $<$ 10$\,$km/s), 
PG$\,$1605$+$072
displays considerable rotation (v$\,$sin$\,$i = 39$\,$km/s, P$<$8.7h) and is 
predicted to evolve into an unusually fast rotating white dwarf.
This nicely confirms a prediction by Kawaler (1999) who deduced a rotation 
velocity of 130km/s from the power spectrum of the pulsations which implies a 
low inclination angle of the rotation axis.}
\KEYWORDS={Stars: atmospheres Stars: abundances Stars: subdwarfs
           Stars: rotation
           Stars: individual: PG1605$+$072, Feige~48, KPD2109$+$4401, 
                              PG1219$+$534
          } 
\SUBMITTED={October 27, 1999}
\printheader
\section{1. INTRODUCTION}

It is now well established that the hot subluminous B stars can be 
identified with models of the extreme Horizontal Branch (EHB) stars (Heber, 
1986, Saffer et al., 1994).

Recently, 
several sdB stars have been found to be pulsating (termed EC14026 stars 
after the prototype, see O'Donoghue et al. 1999 
for a review), 
defining a new instability strip in the HR-diagram. 
The study of these pulsator's 
offers the possibility of applying  
the tools of asteroseismology to investigate the structure of sdB stars.
The existence of pulsating sdB stars was predicted by 
Charpinet et al. (1996), who uncovered an efficient driving mechanism due 
to an opacity bump associated with iron ionization in EHB models. However,
in order to drive the pulsations, iron needed to be enhanced in the appropriate 
subphotospheric layers, possibly due to diffusion. Subsequently, 
Charpinet et al. (1997) confirmed this assumption by detailed
diffusion calculations. Even more 
encouraging was the agreement of the observed and predicted instability 
strip. 

Thirteen pulsating sdB stars are well-studied photometrically (O'Donoghue 
et al. 1999).
A precise knowledge of effective temperature, gravity, element 
abundances and rotation is
a prerequisite for the asteroseismological investigation.

We selected four EC14026 stars for a detailed quantitative spectral analysis: 
PG$\,$1605$+$072 was chosen 
because it has the lowest gravity and, therefore, has probably already 
evolved beyond the extreme horizontal branch phase. It also  
displays the richest 
frequency spectrum amongst the EC$\,$14026 stars ($>$50 periods have been 
identified, Kilkenny al. 1999). Recently, Kawaler (1999) predicted from his 
modelling of the pulsations that PG$\,$1605$+$072 should be rotating. 
PG$\,$1219$+$534 was chosen 
because it has the shortest pulsation periods and has a helium 
abundance larger than most other sdB stars (O'Donoghue et al., 1999).
For Feige$\,$48 and KPD$\,$2109$+$4401 only 4 rsp. 5 frequencies have 
been found so far. Feige$\,$48 is also the coolest of all EC$\,$14026 stars 
known.

\section{2. OBSERVATIONS}

High resolution optical spectra
of the four pulsating sdB stars were obtained with the 
HIRES echelle spectrograph (Vogt et al. 1994) on the Keck 
I telescope on July 20, 1998 using the blue cross 
disperser to cover the full wavelength region between 3700\AA\ and 5200\AA\
at a resolution 0.09\AA.

The spectra are integrated over one 
pulsation cycle or more since the exposure times (600--900s) were long 
compared to the pulsational periods.

The standard data reduction as described by Zuckerman \& Reid (1998)
resulted in spectral orders that have a somewhat wavy 
continuum. In order to remove the waviness we used the spectrum of H1504$+$65 
(a very hot pre-white dwarf devoid of hydrogen and helium) 
which was observed in the same night. Its spectrum has only few weak lines 
of highly ionized metals in the blue (3600--4480\AA) where the strong Balmer 
lines are found in the sdB stars. Therefore we normalized 
individual spectral orders 1 to 20 (3600--4480\AA) of the sdB stars by
dividing through the smoothed spectrum of H1504$+$65. The remaining 
orders were normalized by fitting the continuum with spline functions
(interpolated for orders 26 and 27 which contain H$\beta$). 
Judged from the match of line profiles in 
the overlapping parts of neighboring orders this procedure worked 
extremely well. Atmospheric parameters determined from individual
Balmer lines are found to be consistent with each other except for H$\beta$.
Therefore, we excluded H$\beta$ from the fit procedure. 
Moreover, the resulting T$_{\rm eff}$ 
and log~g are also in excellent agreement with 
those from the fit of a low resolution spectrum of PG$\,$1605$+$072
obtained at the ESO NTT. Details on the analysis of PG$\,$1605$+$072 can be 
found in Heber et al. (1999a).

\section{3. ATMOSPHERIC PARAMETERS}

The simultaneous fitting of Balmer and He line profiles by a grid of 
synthetic spectra (see Saffer et al. 1994) has become the standard 
technique to determine the atmospheric parameters of sdB stars.
The Balmer lines (H$\gamma$ to H$\,$12), He~I 
(4471\AA, 4026\AA, 4922\AA, 
4713\AA, 5016\AA, 5048\AA) and He~II 4686\AA\ lines are fitted to 
derive all three parameters simultaneously.

The analysis is based on
grids of metal line blanketed LTE model atmospheres  
for solar metalicity and Kurucz'
ATLAS6 Opacity Distribution Functions (see Heber et al. 1999b). Synthetic 
spectra are calculated with  
Lemke's LINFOR program (see Moehler et al. 1998).
In addition, a grid of H-He line blanketed, metal free NLTE 
model atmospheres (Napiwotzki 1997), calculated with the ALI 
code of Werner \& Dreizler (1999).

\subsection{PG1605$+$072}

The results (T$_{\rm eff}$=31$\,$900K, log~g=5.29, log(He/H)=-2.54) are in  
agreement with those from low resolution spectra analysed with 
similar models
(Koen et al. 1998) as well as from our own low resolution spectrum for 
PG$\,$1605$+$072.  

\WFigure{1}{\psfig{figure=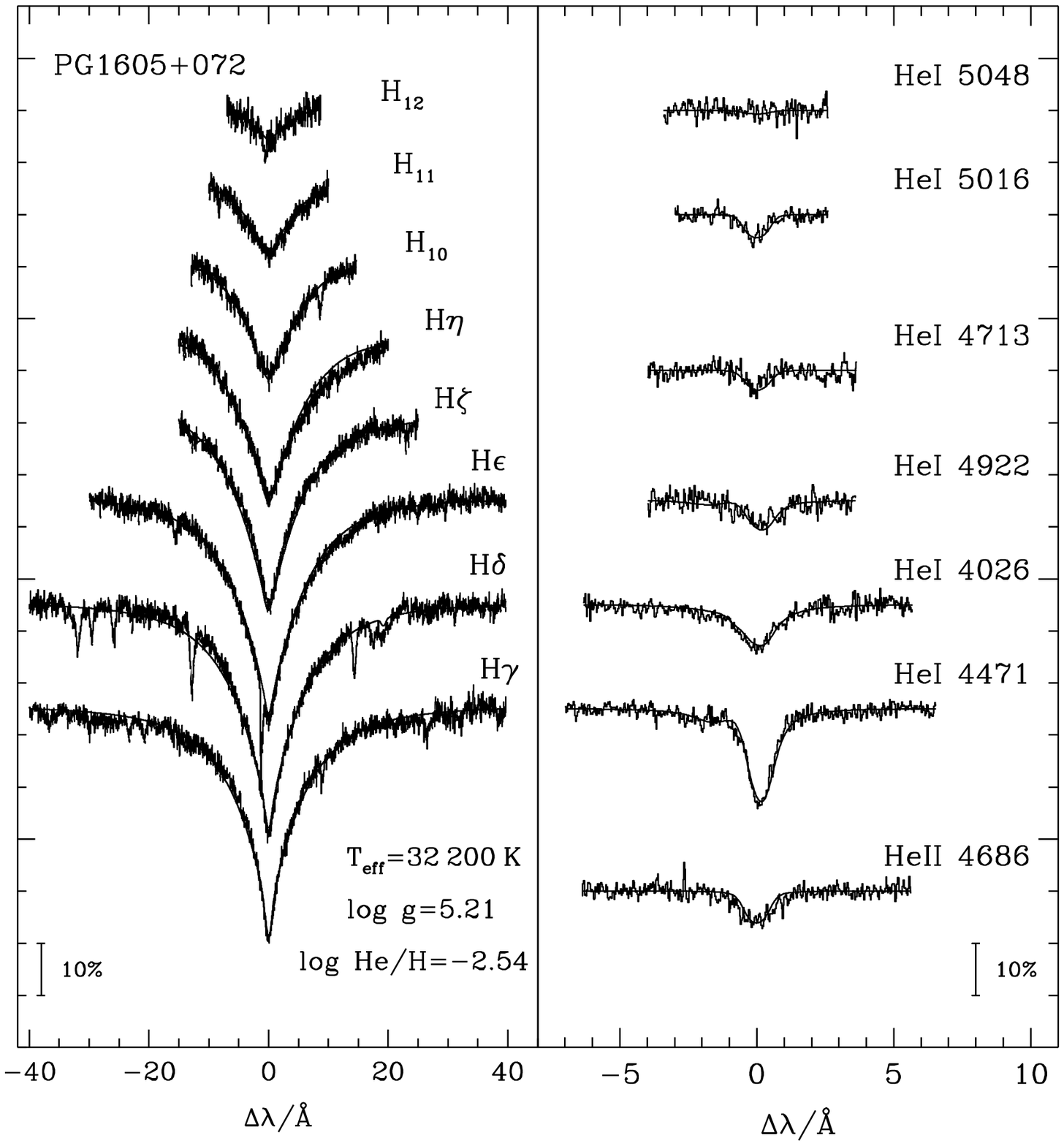,width=15truecm,angle=0,
clip=}}
{Balmer and He line profile fits for PG$\,$1605+072 of the HIRES spectrum
from NLTE model atmospheres.}

Four species are represented by two 
stages of ionization (He~I and He~II, C~II and 
C~III, N~II and N~III, Si~III and Si~IV). 
Since these line ratios 
are
very temperature sensitive at the temperatures in 
question, we alternatively can 
derive T$_{\rm eff}$ and abundances by matching these ionization equilibria.  
Gravity is derived 
subsequently from the Balmer lines by keeping T$_{\rm eff}$ and log(He/H)
 fixed. These
two steps are iterated until consistency is reached. 
C~II is represented by the 4267\AA\ line 
only, which is known to give notoriously too low carbon abundances. Indeed 
the carbon ionization equilibrium can not be matched at any reasonable 
T$_{\rm eff}$. The ionization equilibria of He, N and Si require T$_{\rm eff}$
 to be 
higher than from the Saffer procedure, i.e. 33$\,$200K (He), 33$\,$900K (N) 
and
32$\,$800K (Si). 

Since this difference could be caused by NLTE effects, we repeated the
procedure for T$_{\rm eff}$ and log(He/H) using NLTE models. Alternatively, 
applying Saffer's procedure with the NLTE model grid (see Fig. 1) 
yields T$_{\rm eff}$ almost
identical to that obtained with the LTE grid. 
Evaluating the He ionization equilibrium in NLTE, indeed, results in 
T$_{\rm eff}$
being consistent with that from Saffer's procedure. 
We therefore conclude that 
the higher T$_{\rm eff}$ derived above from the ionization equilibrium in LTE 
is due to NLTE 
effects.
 
However, a systematic difference in log~g persists, the LTE values
being higher by 0.06 -- 0.08 dex than the NLTE results.
Since its origin is obscure, we finally adopted the 
averaged atmospheric parameters: T$_{\rm eff}$=32$\,$300$\pm$300K, 
log~g=5.25$\pm$0.05, 
log(He/H)=-2.53$\pm0.1$. 
Helium is
deficient by a factor of 30 as
is typical for sdB stars.

\subsection{Feige$\,$48}
Since no He~II line can be detected, the helium ionization equilibrium can 
not be evaluated. The procedure of Saffer et al. (1994) results in 
T$_{\rm eff}$=29$\,$400K, log~g=5.51,
log(He/H)=-2.90. Feige$\,$48 has the lowest helium abundance among our 
programme stars. 

\vfill\eject
\subsection{KPD$\,$2109$+$4401}

The helium ionization equilibrium and the Saffer et al. procedure 
give (averaged) parameters T$_{\rm eff}$=31$\,$800K, log~g=5.79,
log(He/H)$=-$2.22 for KPD$\,$2109$+$4401.

\subsection{PG$\,$1219$+$534}

Unlike for PG$\,$1605$+$072, the  helium ionization equilibrium and the 
Saffer et al. procedure give discrepant results: T$_{\rm eff}$=33$\,$200K, 
log~g=5.93, log(He/H)=-1.60 (Saffer et al. procedure, Fig. 2) and 
T$_{\rm eff}$=35$\,$200K, log~g=6.03, log(He/H)=-1.41 (He ionization 
equilibrium, Fig. 3). At the lower T$_{\rm eff}$ the Balmer lines are well 
matched through out the entire profile, whereas for He~II 4686\AA\ there is 
a significant mismatch (see Fig.2). 
At the higher T$_{\rm eff}$ He~II 4686\AA\
is well reproduced, but the Balmer line cores are not reproduced at all 
(see Fig. 3). Despite of its high gravity PG$\,$1219$+$534 has an unusually 
high helium abundance, i.e. it is deficient by a factor of 2 to 5 only.
The line cores of He~I 4026\AA\ and 4471\AA\ cannot be reproduced by 
either model. We conclude that our models do not describe the outermost 
layers of the atmosphere correctly where the cores of the Balmer and He~I 
lines are formed. We point out that PG$\,$1219+534 has the highest helium 
abundance and the shortest pulsation periods, which might affect the outermost
layers.

\WFigure{2}{\psfig{figure=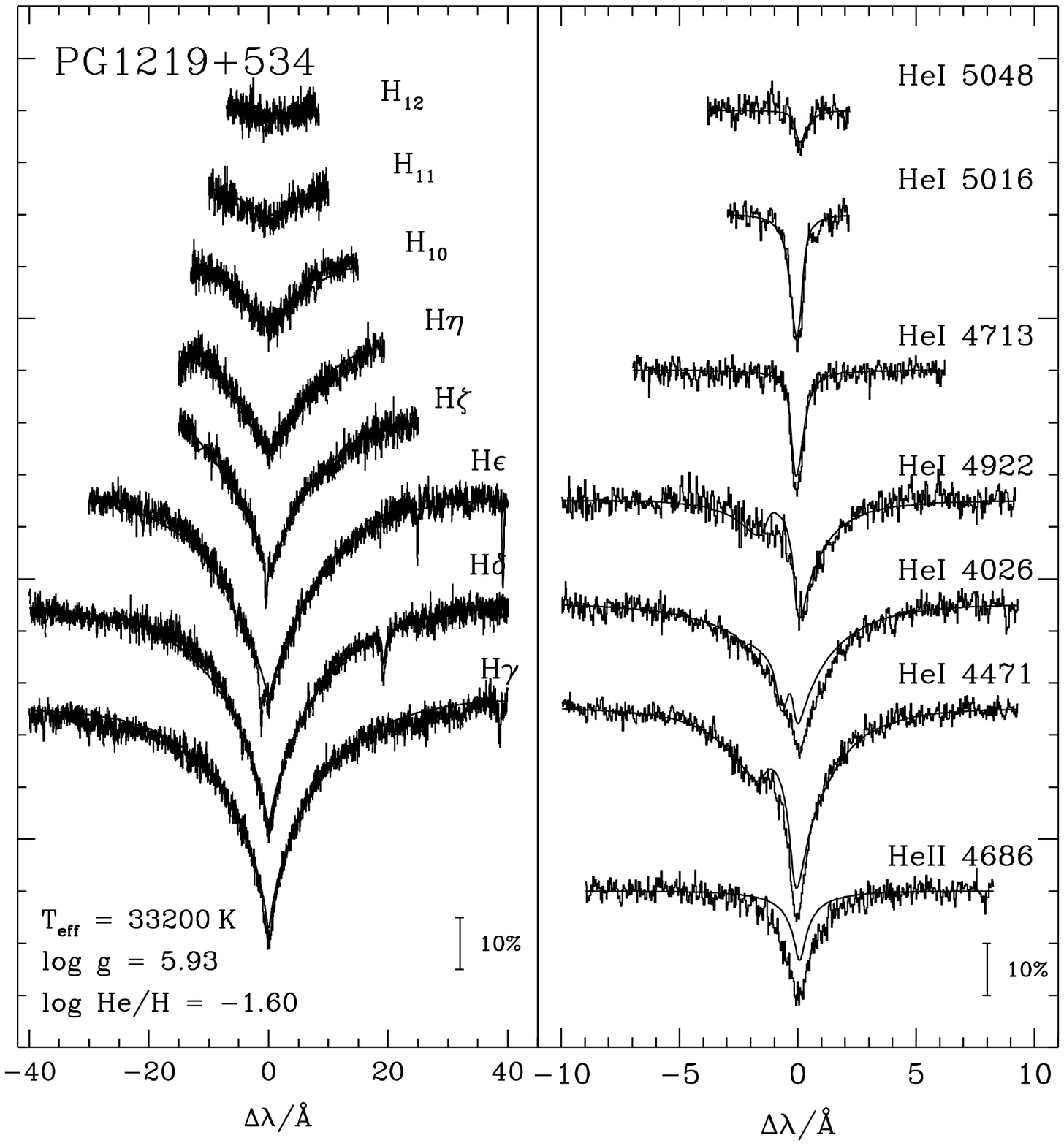,width=15truecm,angle=0,clip=}}
{Balmer and He line profile fits for PG$\,$1219$+$534 of the HIRES spectrum. 
Note the mismatch of the He~II 4686\AA\ line profile and the cores of 
He~I 4026\AA\ and 4471\AA. }

\WFigure{3}{\psfig{figure=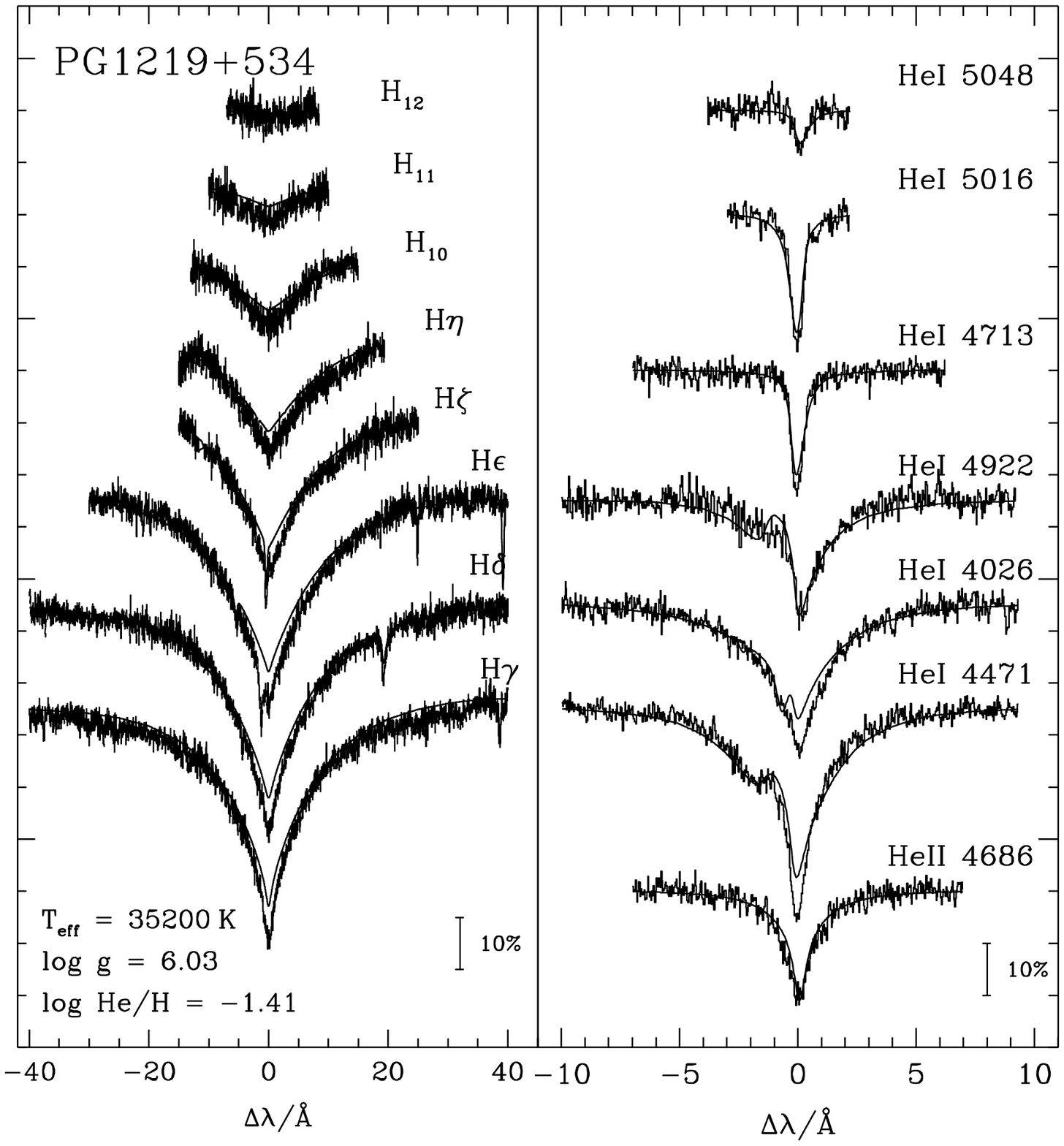,width=15truecm,angle=0,clip=}}
{He line profile fits for PG$\,$1219$+$534 of the HIRES spectrum
to determine T$_{\rm eff}$ and log(He/H) 
simultaneously, log~g is adjusted to match the Balmer line 
wings. Note the mismatch of the cores of the Balmer lines and of He~I 4026\AA\ 
and 4471\AA.}

\section{4. ABUNDANCES}

Weak metal lines are present in the spectra of all programme stars. 
However, the number of detectable lines differs considerably. 
The largest number of metal lines is present in Feige$\,$48 (C, N, O, Ne, Mg, 
Si, Al, S and Fe) and PG$\,$1605$+$072 (which lacks Al and S). 
In PG$\,$1219$+$534 only N, S and Fe are detectable. 

\WFigure{4}{\psfig{figure=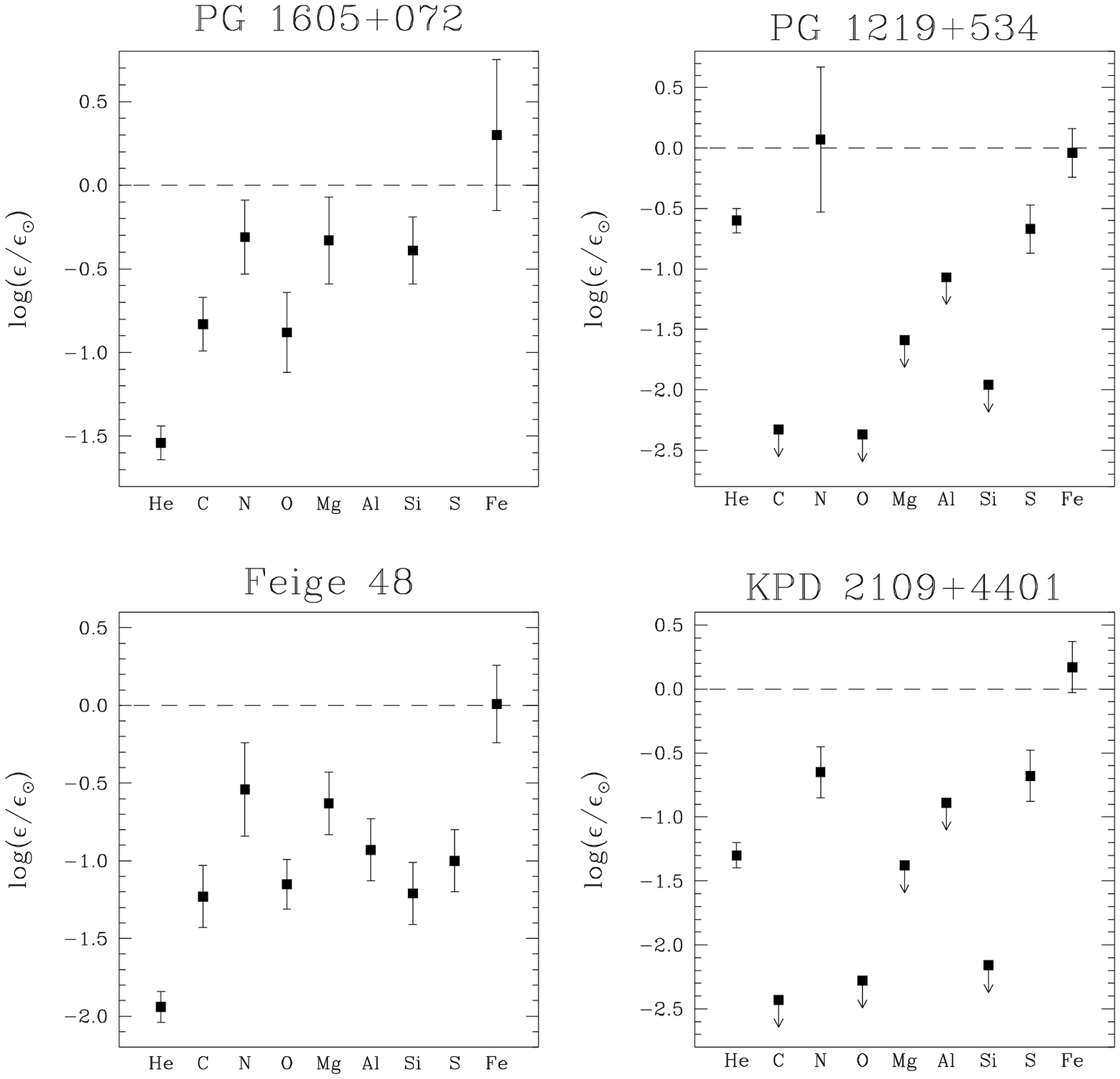,width=15truecm,angle=0,clip=}}
{Abundances of PG$\,$1605$+$072 relative to the sun. 
Upper limits are denoted by arrows. }

The metal lines are sufficiently isolated to derive 
abundances from their equivalent widths except for the crowded region 
from 4635\AA\ to 4660\AA\ in PG$\,$1605$+$072 which we analyse by detailed 
spectrum synthesis.
Results are plotted in Figure 4. Upper limits are shown when no line of a 
species was detectable.
Although several O lines are available in the spectra of PG$\,$1605$+$072 
and Feige$\,$48, it was impossible to determine the 
microturbulent velocity in the usual way, i.e. by minimizing the
slope in a plot of the O abundances versus equivalent widths, due to the 
lack of sufficiently strong lines.
We adopted 5$\pm$5km/s which translates 
into small systematic abundance uncertainties of $\pm$0.05dex for most 
ions. The analysis is done in LTE. 
A temperature uncertainty of $\Delta$T$_{\rm eff}$=1000$\,$K translates 
into abundance uncertainties of less than 0.1$\,$dex. Hence systematic errors 
are smaller for most ions than the statistical errors.

Like helium the metals are deficient with the notable exception of iron, 
which is solar to within the error limits. The high gravity stars 
(KPD$\,$2109$+$4401 and PG$\,$1219$+$534) have considerably lower O and Si 
abundances than the stars of somewhat lower gravity which point to the 
(selective) action of diffusion. It is, however, puzzling that iron is 
solar irrespective of the stellar gravity. UV spectroscopy is required to 
determine more precise iron abundances. We point out that a solar surface 
abundance is in perfect agreement with the diffusion calculations of 
Charpinet et al. (1997).         

\vfill\eject
\section{5. ROTATION VELOCITIES}

The spectral lines of PG$\,$1605$+$072 are considerably broadened,
which we attribute to stellar rotation and derive 
v$\,$sin$\,$i = 39km/s, by fitting the strongest metal lines.
In Figure 5 we compare a section of the spectrum of PG$\,$1605$+$072 to that 
of the Feige$\,$48, which (like PG$\,$1219$+$534 and KPD$\,$2109$+$4401) 
are very sharp-lined 
(v$\,$sin$\,$i$<$8--10km/s).

Assuming a mass of 0.5$\,\rm M_{\odot}$
the radius of R=0.28$\,\rm R_{\odot}$ 
for PG$\,$1605$+$072
follows from the 
gravity. Since sin$\,i$ cannot be constrained 
the corresponding rotation period of PG$\,$1605$+$072 must be smaller than 8.7h.
PG$\,$1605+072 displays the most complex power spectrum with more than 50
frequencies identifiable (Kilkenny et al. 1999), 
39 being bona fide normal pulsation frequencies.

Usually rotation becomes manifest in the power spectrum by 
the characteristic splitting into 
equidistantly spaced multiplet components as is observed e.g. for 
the pre-white dwarf PG$\,$1159$-$035 (rotation period: 1.4$\,d$, Winget 
et al. 1991). Such multiplet's, however, have not been identified for 
PG$\,$1605+072. 
Fast rotation introduces higher order terms that result
in unequally spaced multiplet components.
Recently, Kawaler (1999), was able to identify the five main peaks by 
considering mode trapping and rotational splitting.
He predicted that PG$\,$1605$+$072 should be rapidly rotating (130$\,$km/s). 
The measured v$\,$sin$\,$i=$\,$39$\,$km/s, hence, is a nice confirmation of 
Kawaler's prediction. 
Taken at face value a low inclination angle of 17 degrees results. 

Rotation is interesting also from the point of view of stellar evolution.
PG$\,$1605$+$072 is probably already in a post-EHB phase of evolution
(Kilkenny et al. 1999)
and will evolve directly into a white dwarf, i.e. will shrink 
from its present radius of 0.28$\,\rm R_{\odot}$ to about 0.01$\,\rm R_{\odot}$.
Hence 
PG$\,$1605$+$072 will end 
its life as an unusually fast rotating white dwarf if no loss of angular 
momentum occurs. Isolated white dwarfs, however, are known to be mostly very 
slow rotators (e.g. Heber et al. 1997, Koester et al. 1998).

\WFigure{5}{\psfig{figure=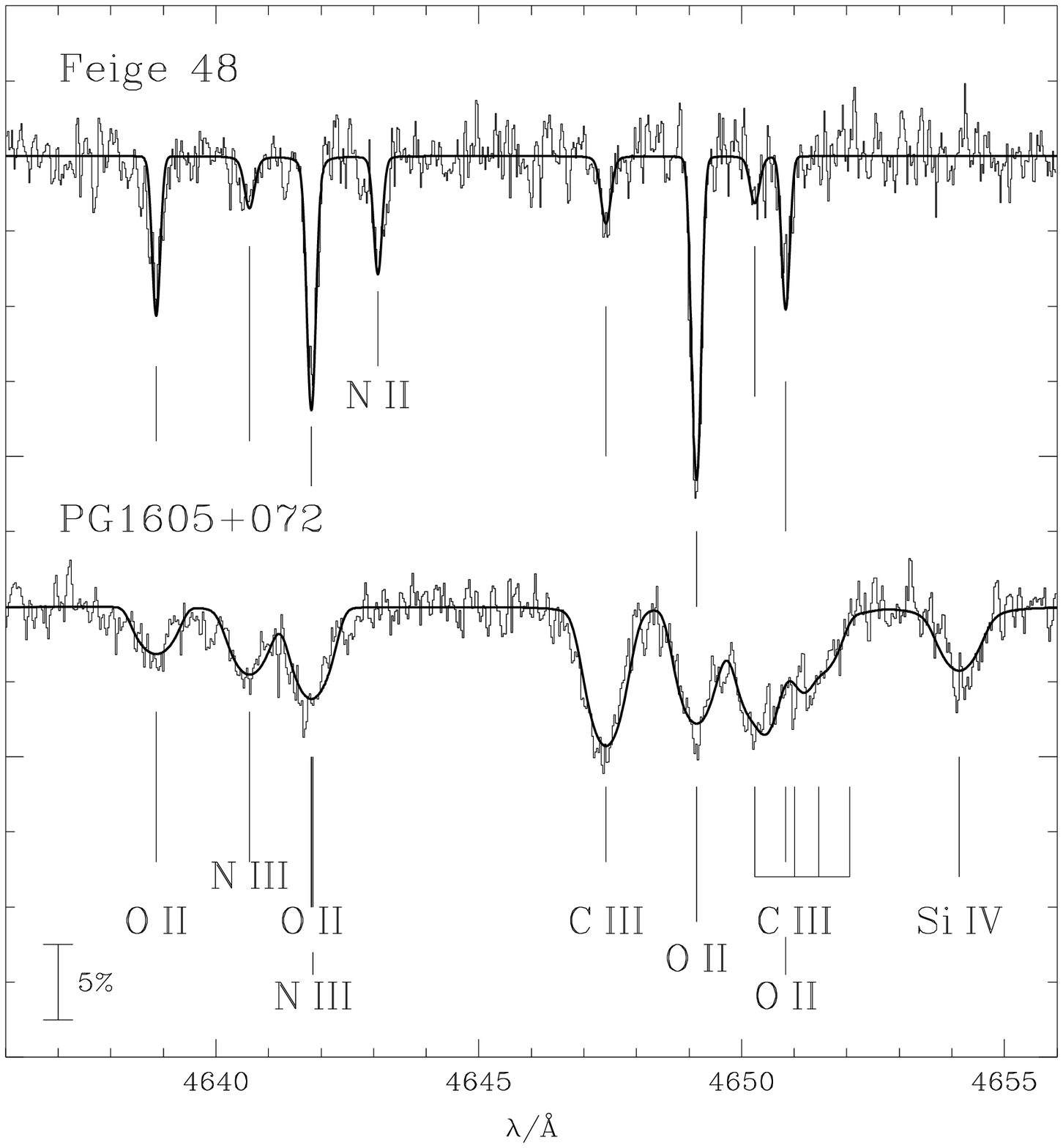,width=12.5truecm,angle=0,clip=}}
{Fit of a section of the metal line spectrum of
PG$\,$1605$+$072, (bottom, v$\,$sini=39$\,$km/s) compared to that of 
Feige$\,$48 (top, no rotation).}

\section{ACKNOWLEDGEMENT.}
U.H. gratefully acknowledges financial support by NATO ARW 
funds.

\vfill\eject
\References
\ref
Charpinet S., Fontaine G., Brassard P., Dorman B. 1996, ApJ 471, 
L103

\ref
Charpinet S., Fontaine G., Brassard P. et al.
1997, ApJ 483, L23 

\ref
Heber U. 1986, A\&A 155, 33

\ref
Heber U., Napiwotzki R., Reid I.N. 1997, A\&A 323, 819

\ref
Heber U., Reid I.N., Werner K. 1999a, A\&A 348,L25

\ref
Heber U., Edelmann H., Lemke M., Napiwotzki R., Engels D., 1999b,
PASPC 169, 551

\ref
Kawaler S. 1999, PASPC 169, 158

\ref
Kilkenny D., et al. 1999, MNRAS 303, 525

\ref
Koen C., O'Donoghue D., Kilkenny D., Stobie R.S. 1998, MNRAS 
           296, 317
\ref
Koester D., Dreizler S., Weidemann V., Allard N.F. 1998, A\&A 338, 612

\ref
Moehler S., Heber U., Lemke M., Napiwotzki R. 1998, A\&A 339, 537

\ref
Napiwotzki R. 1997, A\&A 322, 256

\ref
O'Donoghue D., Koen C., Kilkenny D., Stobie R.S., Lynas-Gray A.E.
           1999, PASPC 169, 149

\ref
Saffer R.A., Bergeron P., Koester D., Liebert J. 1994, ApJ 432, 351

\ref
Vogt S.S., et al. 1994, SPIE 2198, 362

\ref
Werner K. 1991, A\&A 251, 147

\ref
Werner K, Dreizler S. 1999, 
% in Computational Astrophysics, eds. 
%H. Riffert and K. Werner, 
Journal of Computational and Applied 
Mathematics, Elsevier, 109, 65 

\ref
Winget D.E., Nather R.E., Clemens J.C., et al. 1991, ApJ 378, 326 

\ref
Zuckerman B., Reid I.N. 1998, ApJ 505, L143

\bye